\documentclass[a4paper,10pt,twocolumn,twoside,journal]{article}

\usepackage{geometry}
\advance\oddsidemargin by -1in
\advance\evensidemargin by -1in
\advance\headsep by 2.8mm
\usepackage{cite}
\usepackage{hyperref}
\usepackage{url}
\usepackage{authblk}
\usepackage{mathptmx}
\usepackage{braket}
\usepackage{booktabs}
\usepackage{amsthm}
\usepackage{amssymb,amsmath,mdwlist,mathtools}
\usepackage{multirow}
\usepackage{multicol}
\usepackage{hhline}
\usepackage{algorithm}
\usepackage{algorithmicx}
\usepackage{algpseudocode}
\usepackage{graphicx}
\usepackage{caption}
\usepackage{subcaption}
\usepackage{adjustbox}
\usepackage{textcomp}
\usepackage{wrapfig}
\def\BibTeX{{\rm B\kern-.05em{\sc i\kern-.025em b}\kern-.08em
    T\kern-.1667em\lower.7ex\hbox{E}\kern-.125emX}}

\usepackage[normalem]{ulem}

\title{A Modular Quantum Compilation Framework\\for Distributed Quantum Computing}

\author[1,*]{Davide~Ferrari}
\author[2]{Stefano~Carretta}
\author[1]{Michele~Amoretti}
\affil[1]{\small \textit{Quantum Software Laboratory}, Department of Engineering and Architecture, University of Parma, Parma, 43124 Italy (e-mail: \href{mailto:michele.amoretti@unipr.it}{michele.amoretti@unipr.it}, \href{mailto:davide.ferrari1@unipr.it}{davide.ferrari1@unipr.it}. Web: \href{http://www.qis.unipr.it/quantumsoftware.html}{http://www.qis.unipr.it/quantumsoftware.html}}
\affil[2]{\small Department of Mathematical, Physical and Computer Sciences, University of Parma, I-43124 Parma, Italy and INFN–Sezione di Milano-Bicocca, gruppo collegato di Parma, 43124 Parma, Italy}
\affil[*]{\small Corresponding author: Davide Ferrari, davide.ferrari1@unipr.it.}

\date{}

\begin{document}

\maketitle

\begin{abstract}\textbf{
For most practical applications, quantum algorithms require large resources in terms of qubit number, much larger than those available with current NISQ processors. With the network and communication functionalities provided by the Quantum Internet, Distributed Quantum Computing (DQC) is considered as a scalable approach for increasing the number of available qubits for computational tasks. For DQC to be effective and efficient, a quantum compiler must find the best partitioning for the quantum algorithm and then perform smart remote operation scheduling to optimize EPR pair consumption. At the same time, the quantum compiler should also find the best local transformation for each partition. In this paper we present a modular quantum compilation framework for DQC that takes into account both network and device constraints and characteristics. We implemented and tested a quantum compiler based on the proposed framework with some circuits of interest, such as the VQE and QFT ones, considering different network topologies, with quantum processors characterized by heavy hexagon coupling maps. We also devised a strategy for remote scheduling that can exploit both TeleGate and TeleData operations and tested the impact of using either only TeleGates or both. The evaluation results show that TeleData operations may have a positive impact on the number of consumed EPR pairs, while choosing a more connected network topology helps reduce the number of layers dedicated to remote operations.}
\end{abstract}


\maketitle

\section{Introduction}
\label{sec:introduction}

Noisy Intermediate-Scale Quantum (NISQ) processors are characterized by few hundreds of quantum bits (qubits) with non-uniform quality and highly constrained physical connectivity. Hence, the growing demand for large-scale quantum computers is motivating research on Distributed Quantum Computing (DQC) architectures \cite{DQCSurvey2022} as a scalable approach for increasing the number of available qubits for computational tasks, and experimental efforts have demonstrated some of the building blocks for such a design \cite{VanDev2016}. Indeed, with the network and communications functionalities provided by the \textit{Quantum Internet} \cite{PirBra2016,Gib2016,DurLamHeu2017,Sim2017,Wehner2018,Zomo2018,CalCacBia2018,CalChaCuo2020,GyoImr2020,GyoImr2020-2,AmoCar2020}, remote quantum processing units (QPUs) can communicate and cooperate for executing computational tasks that each NISQ device cannot handle by itself.

\begin{figure}
    \centering
    \begin{minipage}[c]{.7\linewidth}
        \includegraphics[width=1.\textwidth]{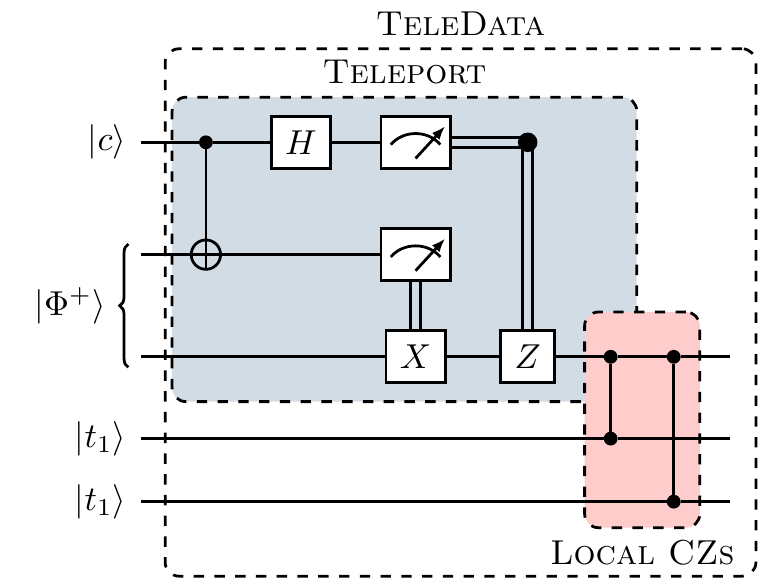}
        \subcaption{}
        \label{fig:tele_data}
    \end{minipage}
    \hfil
    \begin{minipage}[c]{.7\linewidth}
        \includegraphics[width=1.\textwidth]{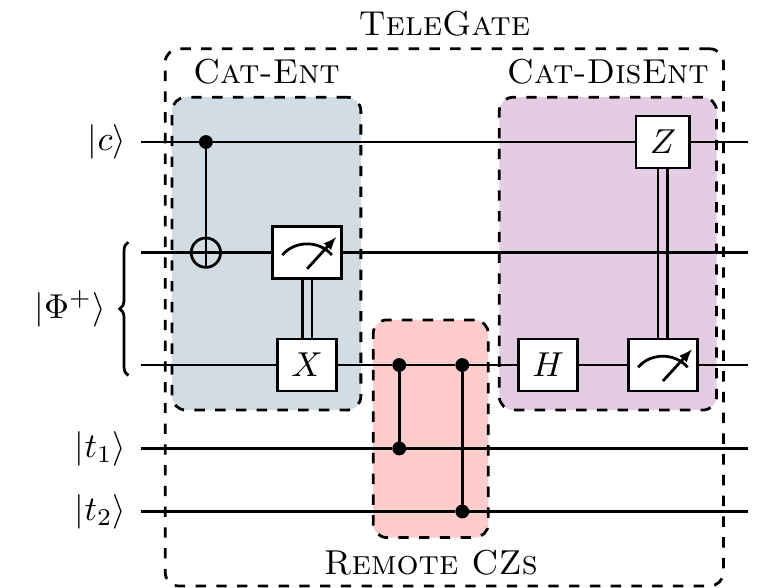}
        \subcaption{}
        \label{fig:tele_gate}
    \end{minipage}
    \caption{\textbf{(\ref{fig:tele_data})} Circuit representation of \texttt{TeleData} by means of the \texttt{Teleport} primitive, which allows one to move the quantum state of a data qubit $\ket{c}$ to the second half of an entangled pair. While both the sate of the entangled pair and the original qubit $\ket{c}$ are now lost, multiple \texttt{CZ} acting on the teleported qubit can then be executed. \textbf{(\ref{fig:tele_gate})} Circuit representation of \texttt{TeleGate} by means of \texttt{Cat-Ent} and \texttt{Cat-DisEnt} primitives. After the \texttt{Cat-Ent} operation, the second half of the entangled pair acts as a shared copy (not an actual copy, due to the no cloning theorem) of the original $\ket{c}$ control qubit. Multiple remote \texttt{CZ} with same control qubit and different target can be executed between \texttt{Cat-Ent} and \texttt{Cat-DisEnt}. It is worth noting that, between \texttt{Cat-Ent} and \texttt{Cat-DisEnt}, the $\ket{c}$ control qubit is entangled with its shared copy and cannot be targeted by other gates.}
    \label{fig:remote_operations}
    \hrulefill
\end{figure}

In general, when moving from local to distributed quantum computing one faces two main challenges, namely, quantum algorithm partitioning and execution management \cite{DQCSurvey2022}. To partition a monolithic quantum algorithm, a \textit{quantum compiler} must be used to find the best breakdown, i.e., the one that minimizes the number of gates that are applied to qubits stored at different devices. Such \textit{remote gates} can be implemented by means of three communication primitives that we denote as \texttt{Teleport}~\cite{CacCalVan2020} (quantum state teleportation), \texttt{Cat-Ent} (cat-entanglement) and \texttt{Cat-DisEnt} (cat-disentanglement)~\cite{Yimsiriwattana2005}. These primitives require that an entangled state is consumed and a new one must be distributed between the remote processors through the quantum link before another inter-processor operation can be executed. Through this primitives one can perform two types of remote operations, namely \texttt{TeleData} and \texttt{TeleGate}~\cite{VanNem2006, VanMun2008, VanDev2016}, as shown in Fig.~\ref{fig:remote_operations}.
The literature on quantum compilers~\cite{AndresMartinez2019,Sundaram2021,SunGupRam2022,Daei2020,Dava2020,Dadkhah2021,Nikahd2021,CuoCalKrs2023,Ovide2023,Baker2020,FerAmo2021} focuses on qubit assignment and remote gate scheduling, while pay less attention to the integration of local routing.

\begin{figure}[!ht]
    \centering
    \includegraphics[width=.45\textwidth]{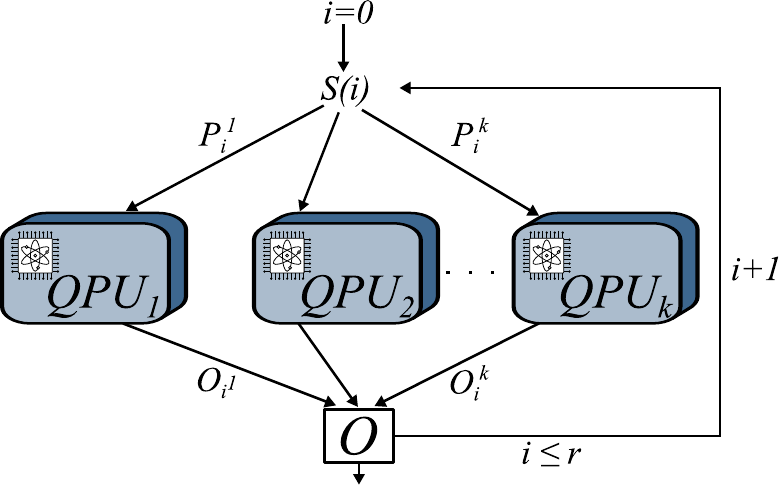}
    \caption{Execution of multiple quantum circuit instances with $k$ QPUs. For each execution round $i$, a schedule $S(i)$ maps some quantum circuit instances to the quantum network -- each QPU receiving a quantum circuit $P_i^j$ that is either a monolithic one or a sub-circuit of a monolithic one. The classical outputs are accumulated into an output vector $O$.}
    \label{fig:exec_manage}
    \hrulefill
\end{figure}

Regarding execution management, in general, given a collection $\mathcal{P}$ of quantum circuit instances to be executed, this collection should be divided into non-overlapping subsets $\mathcal{P}_i$, such that $\mathcal{P} = \cup_i \mathcal{P}_i$. One after the other, each subset must be assigned to the available QPUs. In other words, for each execution round $i$, there exists a schedule $S(i)$ that maps some quantum circuit instances to the quantum network. If DQC is supported, some quantum circuit instances may be split into sub-circuit instances, each one to be assigned to a different QPU, as illustrated in Figure~\ref{fig:exec_manage}).

In this work, we focus on the first challenge, i.e., quantum algorithm partitioning. We present a modular quantum compilation framework for DQC that, for the first time, takes into account both network and device constraints and characteristics.
We illustrate the experimental evaluation of a quantum compiler based on the proposed framework, using some circuits of interest (VQE, QFT, graph state preparation) and different network topologies, with quantum processors characterized by heavy hexagon coupling maps. The heavy-hexagon topology has been chosen by IBM~\cite{IBMTopologies} for its scalability and performance, offering reduced error-rates while affording the opportunity to explore error correcting codes~\cite{Takita2017,Sundaresan2020,Córcoles2015,Chamberland2020,CuoCalKrs2023}. We also devised a strategy for remote scheduling that can exploit both \texttt{TeleGate} and \texttt{TeleData} operations, and tested the impact of using either only \texttt{TeleGate}s or both.
The evaluation results show that \texttt{TeleData} operations may have a positive impact on the number of consumed EPR pairs, while choosing a more connected network topology helps reduce the number of layers dedicated to remote operations.

The paper is organized as follows. In Section \ref{sec:related}, related works on quantum compiling for DQC are discussed. In Section \ref{sec:compiler_framework}, the proposed modular quantum compilation framework is illustrated in detail. In Section \ref{sec:evaluation}, the experimental evaluation of a Python-based implementation of the compiler is presented. Finally, Section \ref{sec:conclusion} concludes the paper with a discussion of open questions and future work.

\section{Related Work}
\label{sec:related}

Quantum compilation for DQC is characterized by two fundamental steps, \textit{qubit assignment} and \textit{remote gate scheduling}. In DQC, qubit assignment is generally tackled as a partitioning problem. Specifically, for a given set of virtual qubits, one needs to choose a partition that maps sub-sets of logical qubits to processors, while minimizing the number of required interactions among different sub-sets. The main goal is to minimize the number of consumed \textit{ebits} -- i.e., EPR pairs shared between QPUs --, as it is the main bottleneck to distributed quantum computation.

Andr\'es-Mart\'inez and Heunen~\cite{AndresMartinez2019} use \textit{cat-entanglement}~\cite{Yimsiriwattana2005} to implement remote quantum gates. The chosen gate set contains every one-qubit gate and two single two-qubit gates, namely the \texttt{CNOT} and the \texttt{CZ} gate (i.e., the controlled version of the \texttt{Z} gate). The authors consider no restriction on the ebit connectivity between QPUs. Then, they reduce the problem of distributing a circuit across multiple QPUs to hypergraph partitioning. The proposed approach is evaluated against five quantum circuits, including QFT. The proposed solution has some drawbacks, in particular that there is no way to customize the number of communication qubits of each QPU.

Sundaram et al.~\cite{Sundaram2021} present a two-step solution, where the first step is qubit assignment. Circuits are represented as edge-weighted graphs with qubits as vertices. The edge weights correspond to an estimation for the number of cat-entanglement operations. The problem is then solved as a minimum k-cut, where partitions have roughly the same size. The second step is finding the smallest set of cat-entanglement operations that will enable the execution of all \texttt{TeleGate}s. The authors state that, in a special setting, this problem can be reduced to a vertex-cover problem, allowing for a polynomial-time optimal solution based on integer linear programming. They also provide a $O(\log n)$-approximate solution, where $n$ is the total number of global gates, for a generalized setting by means of greedy search algorithm. In~\cite{SunGupRam2022}, the same authors extend their approach to the case of an arbitrary-topology network of heterogeneous quantum computers by means of a Tabu search algorithm.

In~\cite{Daei2020}, by Daei et al., the circuit becomes an undirected graph with qubits as vertices, while edge weights correspond to the number of two-qubit gates between them. Then, the graph is partitioned using the Kernighan-Lin (K-L) algorithm for VLSI design~\cite{KerLin1970}, so that the number of edges between partitions is minimized. Finally, each graph partition is converted to a quantum circuit.

In~\cite{Dava2020}, the authors represent circuits as bipartite graphs with two sets of vertices -- one set for the qubits and one for the gates -- and edges to encode dependencies of qubits and gates. Then, for the qubit assignment problem, they propose a partitioning algorithm via dynamic programming to minimize the number of \texttt{TeleData} operations.

Dadkhah et al.~\cite{Dadkhah2021} propose a heuristic approach to replace the equivalent circuits in the initial quantum circuit. Then, they use a genetic algorithm to partition the placement of qubits so that the number of teleportations could be optimized for the communications of a DQC.

Nikahd et al.~\cite{Nikahd2021} exploit a minimum k-cut partitioning algorithm formulated as an ILP optimization problem, to minimize the number of remote interactions. They use a moving window and apply the partitioning algorithm to small sections of the circuit, thus the partition may change with the moving window by means of \texttt{TeleData} operations.

Cuomo et al. in~\cite{CuoCalKrs2023} model the compilation problem with an Integer Linear Programming formulation. The formulation is inspired to the vast theory on dynamic network problems. Authors manage to define the problem as a special case of \textit{quickest multi-commodity flow}. Such a result allows to perform optimization by means of techniques coming from the literature, such as a \textit{time-expanded} representation of the distributed architecture.

Ovide et al.~\cite{Ovide2023}, investigate the performance of the qubit assignment strategy proposed by Baker et al.~\cite{Baker2020} on some circuits of interests, under the assumption of local and network all-to-all connectivity. In~\cite{Baker2020}, qubit assignment is treated as a graph partitioning problem, under the assumption that a SWAP operation primitive exists to exchange data qubits between different QPUs. I.e., it is not required to check if there are available free data qubits on the QPUs. Ovide et al. show that, in general, the wider the circuit, the higher the number of remote operations, although it highly depends on the specific circuit to be compiled.

\section{Modular Quantum Compilation Framework}
\label{sec:compiler_framework}

\begin{figure*}[!ht]
    \centering
    \includegraphics[width=.9\textwidth]{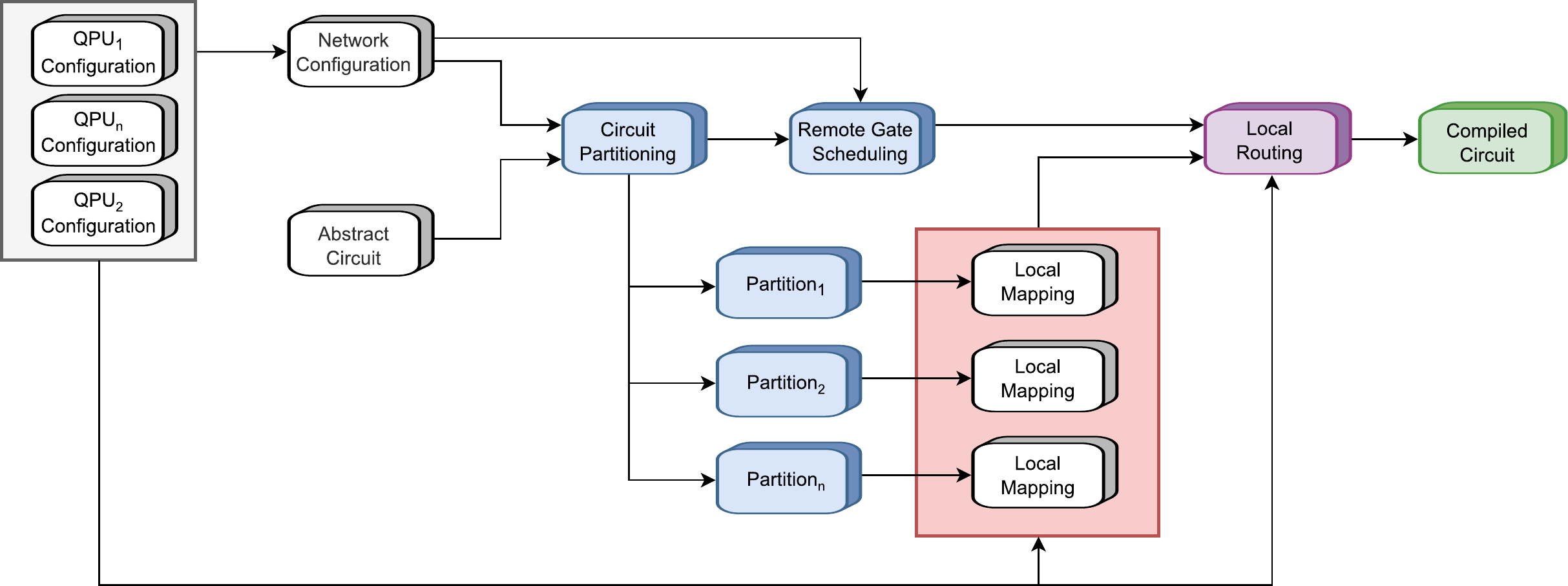}
    \caption{Workflow of the proposed modular quantum compilation framework for DQC architectures.}
    \label{fig:framework}
    \hrulefill
\end{figure*}

As mentioned in Section~\ref{sec:introduction}, there is a lack of a modular framework for compiling quantum circuits to DQC architectures. Such a framework should be circuit agnostic, i.e., able to compile any circuit to any suitable DQC architecture. Moreover, this framework should bridge the gap between local compilation and compilation for DQC. Current proposals from the literature tackle the problems of qubit assignment and remote gate scheduling but do not take into account the local connectivity of each QPU.
Our proposal for a general-purpose quantum compilation framework is shown in Fig.~\ref{fig:framework}.

\begin{figure}[!ht]
    \centering
    \includegraphics[width=.8\linewidth]{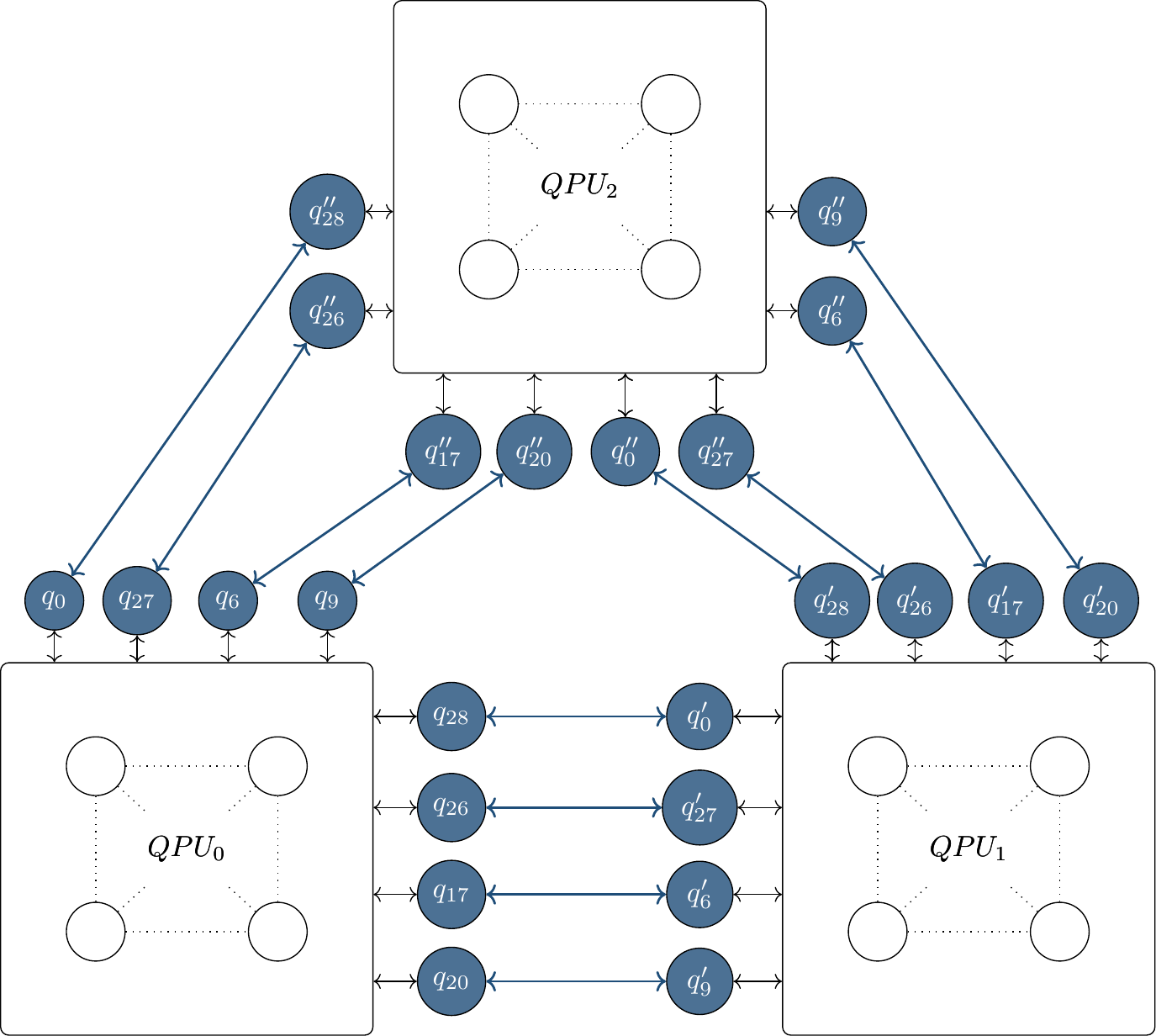}
    \caption{DQC architecture comprising 3 QPUs as shown in Fig.~\ref{fig:input_device}. Each QPU is connected to the others and each QPU supports up to 4 \textit{communication qubits} per connection.}
    \label{fig:net_3}
    \hrulefill
\end{figure}

\begin{figure}[!ht]
    \centering
    \includegraphics[width=.9\linewidth]{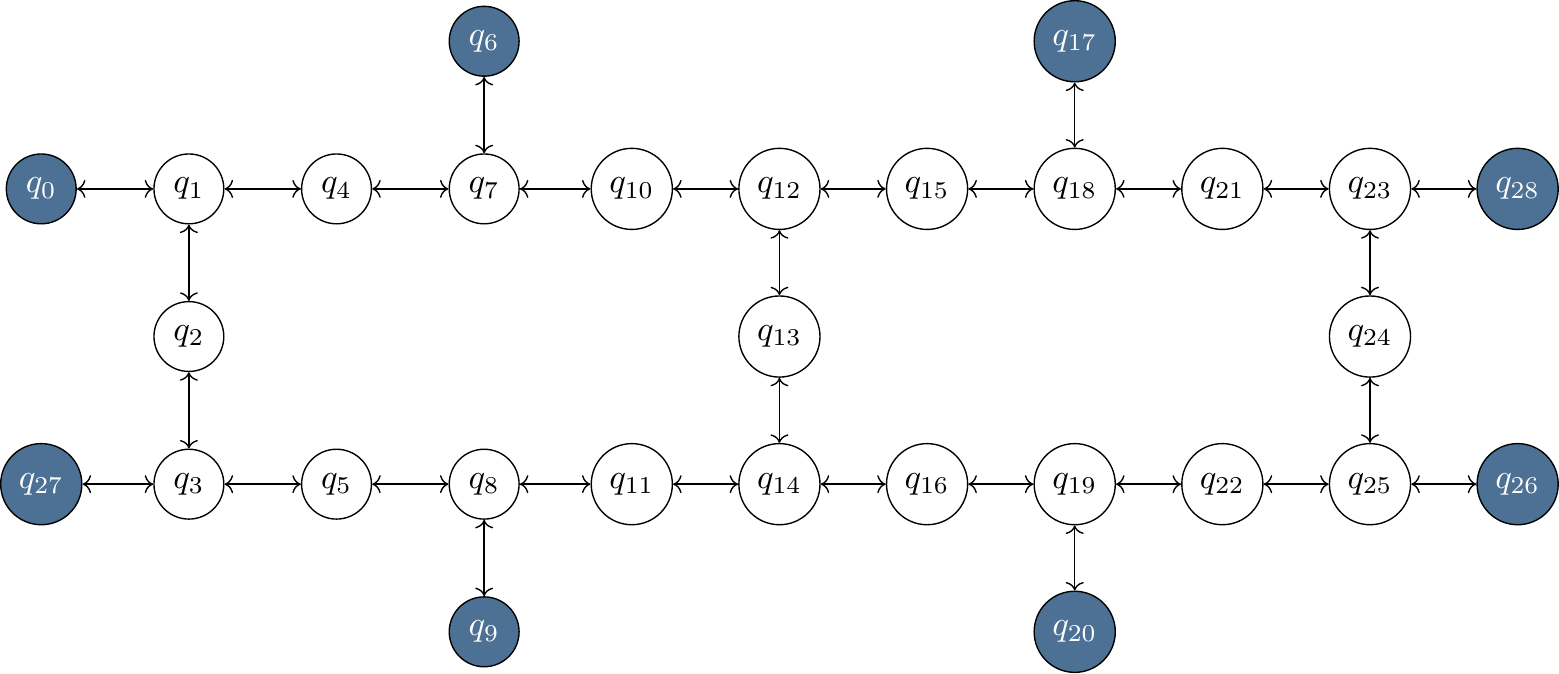}
    \caption{QPU configuration with 20 \textit{data qubits} and 8 \textit{communication qubits}, inspired by IBM's heavy hexagon devices~\cite{IBMQ-backends}.}
    \label{fig:input_device}
    \hrulefill
\end{figure}

The proposed quantum compilation framework takes as input a quantum circuit and a network configuration. As depicted in Fig.~\ref{fig:net_3}, the network configuration describes how QPUs are connected into the target DQC architecture, including quantum channels capacity, i.e., the number of communication qubits for each channel. The network configuration should include descriptions of the internal configurations of the QPUs, i.e., the coupling map and the set of available data qubits and communication qubits. The coupling map is a directed graph where each vertex corresponds to a qubit and directed edges determine the possibility of executing two-qubit gates between the connected qubits\footnote{Specifically, the source and destination vertexes of a directed edge can be the control and target qubit respectively of a two-qubit gate. An edge could be undirected, meaning that both qubits can act as control or target.}. Fig.~\ref{fig:input_device} shows an example of coupling map with 20 data qubits and 8 communication qubits, highlighted in blue.

Having these inputs, the first step in the framework regards the qubit assignment, as described in Sec.~\ref{sec:assignment}. Once a good qubit assignment has been found, the compiler proceeds to schedule remote gates accordingly and computes the local mapping of qubits assigned for each QPU, as detailed in Sec.~\ref{sec:scheduling}. Finally, local routing must be performed while taking into account the previously scheduled remote gates, as explained in Sec.~\ref{sec:routing}.

The output of the framework is of course a compiled circuit. This circuit is a unique object containing all properly scheduled local and remote gates.

\subsection{Qubit Assignment}
\label{sec:assignment}

\begin{figure}[!h]
    \centering
    \begin{minipage}[c]{.7\linewidth}
        \includegraphics[width=1.\textwidth]{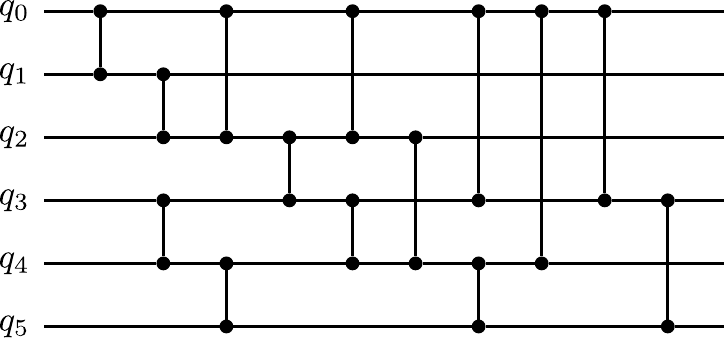}
        \subcaption{}
        \label{fig:circuit}
    \end{minipage}
    \hfil
    \begin{minipage}[c]{.7\linewidth}
        \includegraphics[width=1.\textwidth]{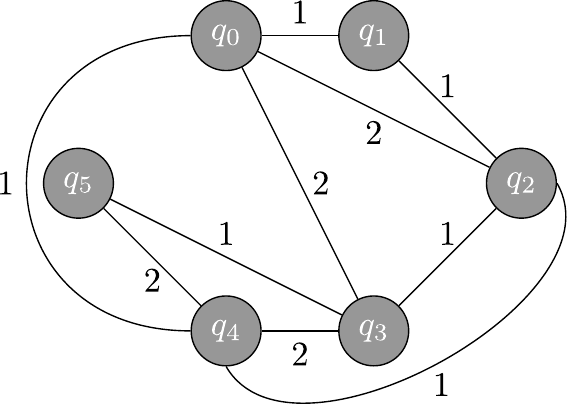}
        \subcaption{}
        \label{fig:circuit_graph}
    \end{minipage}
    \caption{The quantum circuit in \textbf{(\ref{fig:circuit})} can be represented as the weighted graph in \textbf{(\ref{fig:circuit_graph})}.}
    \label{fig:weighted_graph}
    \hrulefill
\end{figure}

As mentioned in Sec.~\ref{sec:introduction}, the goal is to partition the circuit in order to minimize the communication cost, i.e., the number of remote operations and consequently the number of consumed EPR pairs. To this aim, a quantum circuit $qc$ can be represented as an undirected weighted graph $G_{qc}(V,E)$, as shown in Fig.~\ref{fig:weighted_graph}, where each edge $e \in E$ has weight $W(e) \in \mathbb{N}$. The set of vertices $V$ corresponds to the qubits in $qc$ and the weight of each edge is equal to the number of two-qubit gates between the corresponding qubits.

\begin{figure*}[!h]
    \centering
    \begin{minipage}[c]{.4\linewidth}
        \includegraphics[width=1.\textwidth]{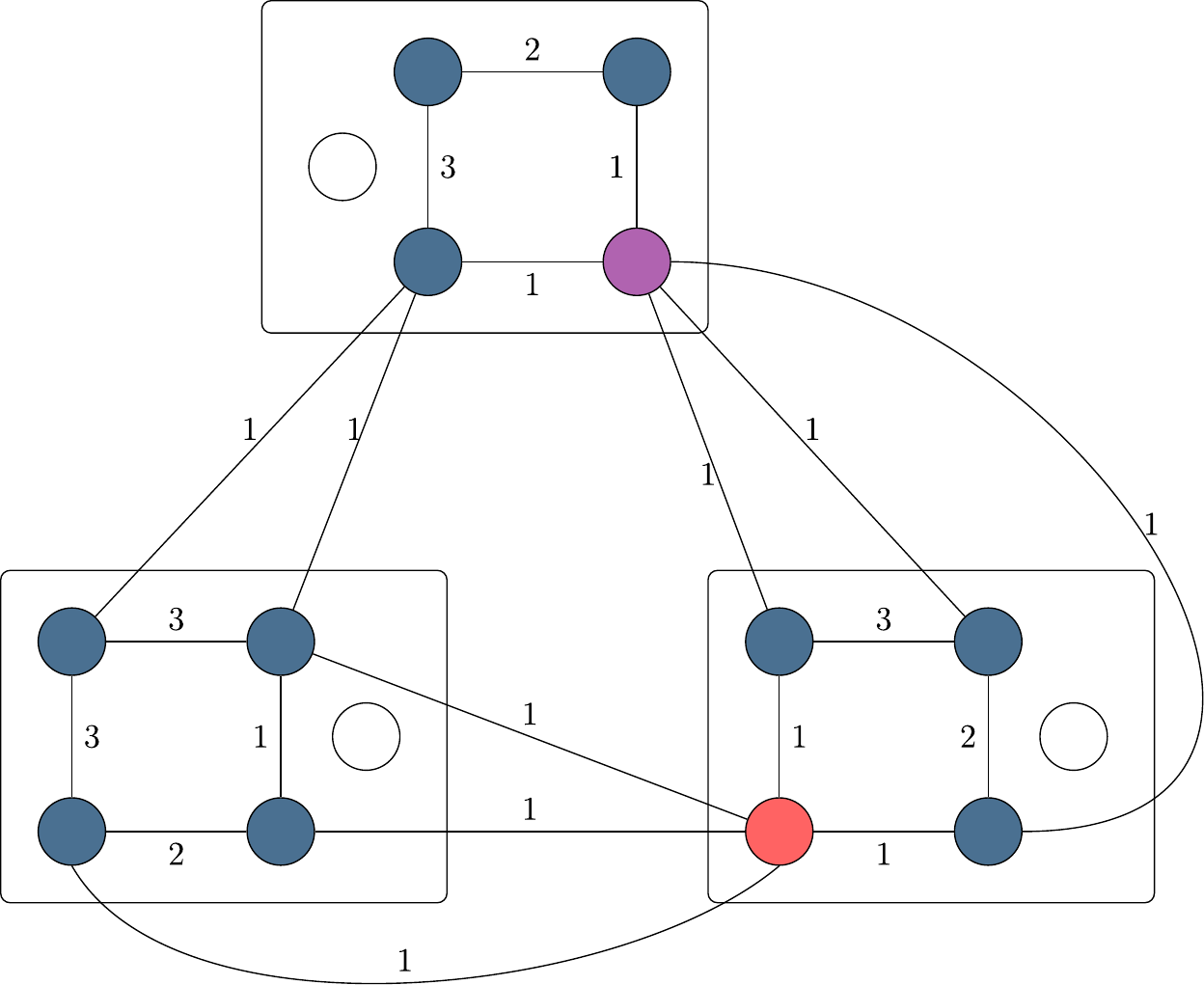}
        \subcaption{}
        \label{fig:partitioning_1}
    \end{minipage}
    \hfil
    \begin{minipage}[c]{.4\linewidth}
        \includegraphics[width=1.\textwidth]{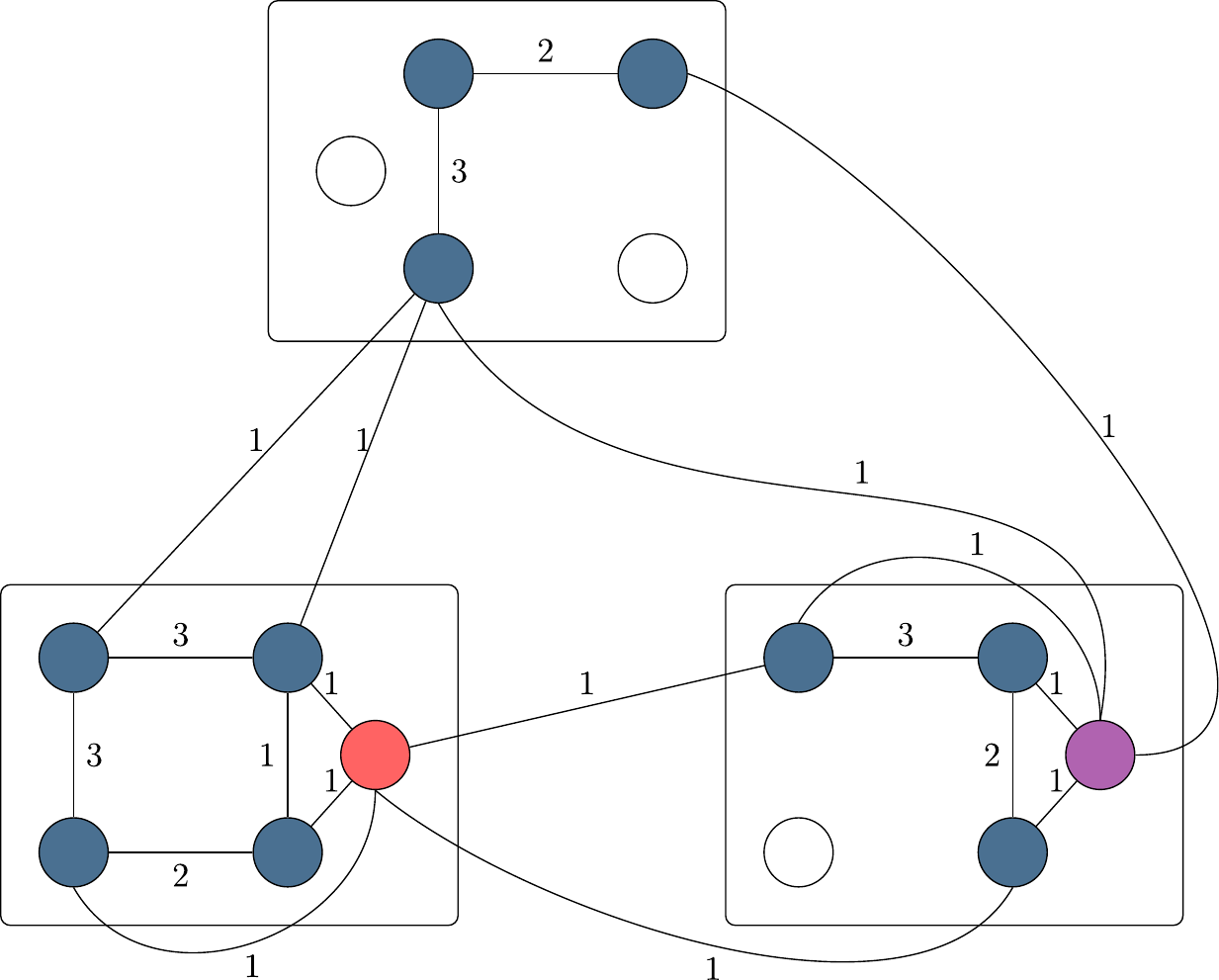}
        \subcaption{}
        \label{fig:partitioning_2}
    \end{minipage}
    \caption{\textbf{(a)} Example of initial graph partitioning. There are three partitions, each holding 4 qubits, and 8 edges between different partitions. The total communication cost is equal to 8. White nodes represent available qubits that have not been utilized. \textbf{(b)} Graph partitioning where qubits have been moved to achieve a better solution. The total communication cost is now equal to 6.}
    \label{fig:partitioning}
    \hrulefill
\end{figure*}

The qubit assignment problem can then be treated as a graph partitioning problem where the objective is to compute a k-way partitioning such that the sum of edges' weights that straddle different partitions is minimized. Given $k$ available QPUs, the result of k-way partitioning are $k$ roughly equally sized circuit partitions. There are several algorithms available that can efficiently find a solution. In this work we used METIS’s multilevel k-way partitioning\footnote{\href{https://github.com/KarypisLab/METIS}{METIS GitHub repository.}}. This approach is preliminary and not optimal, as the QPUs would probably be underutilized and the circuit's qubits unnecessarily scattered through all QPUs. In fact, for each partition we check if moving a qubit to another partition would benefit the overall communication cost. Between all the useful moves found, we choose the best and iteratively continue to search for possible movements, until either all qubits have been moved one time or no more good movements can be found. An example of improvement from the initial solution is depicted in Fig.~\ref{fig:partitioning}

\subsection{Remote Gate Scheduling}
\label{sec:scheduling}

\begin{figure}
    \centering
    \begin{minipage}[c]{.52\linewidth}
        \includegraphics[width=1.\textwidth]{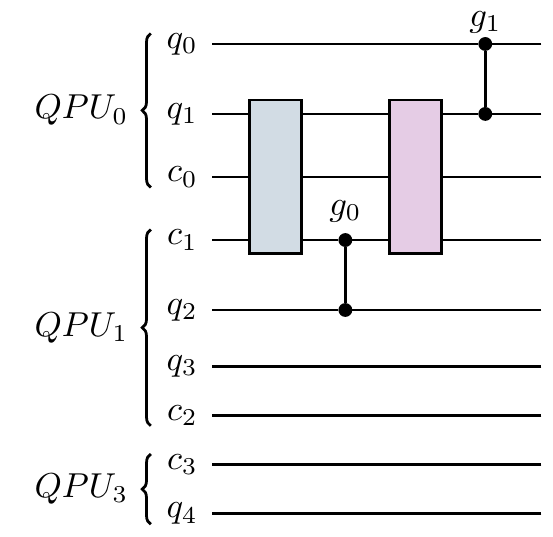}
        \subcaption{}
        \label{fig:one_mig}
    \end{minipage}
    \hfil
    \begin{minipage}[c]{.47\linewidth}
        \includegraphics[width=1.\textwidth]{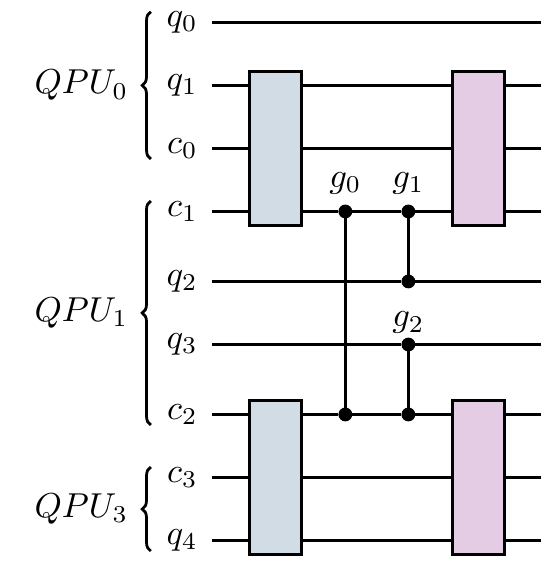}
        \subcaption{}
        \label{fig:two_mig}
    \end{minipage}
    \caption{Gates can be covered by either migrating one qubit to the other qubit's QPU or both to a different one. This concept is valid for both \texttt{TeleGate} and \texttt{TeleData} operations. \textbf{(a)} Gate $g_0$ is covered by sharing qubit $q_1$ with $QPU_1$ using one \texttt{TeleGate}. \textbf{(b)} Qubits $q_1$ and $q_4$ are shared with $QPU_2$ using two \texttt{TeleGate}s, gates $g_0$, $g_1$ and $g_2$ are consequently covered.}
    \label{fig:migration}
    \hrulefill
\end{figure}

We designed a compilation pass to schedule remote gates for DQC architectures to investigate the impact of using both \texttt{TeleData} and \texttt{TeleGate} operations. The main strategy, described in Alg.~\ref{alg:schedule}, requires three input items: the quantum circuit to distribute, the configuration of the network onto which such circuit will be executed and a suitable qubit assignment, as computed by a previous pass in the compilation framework. The pass scans the quantum circuit gate by gate and stops when it encounters a gate that, based on the current partitioning, involves qubits on different QPUs. The pass then searches for feasible \texttt{TeleData} operations that could be covered\footnote{Here, ``to cover'' means to make the gate executable.} by teleporting one or both qubits on a common QPU. \texttt{TeleData} operations are selected by taking into account the memory capacity of each QPU, all the while making sure that no data qubits storing valuable information gets overwritten by a teleportation. Finally, each possible \texttt{TeleData} is assigned a cost, which is given by Eq.~\ref{eq:cover_cost}:

\begin{equation}
    \label{eq:cover_cost}
    \frac{n_{EPR}}{n_{cov}} \, \frac{delay}{\bar{d_{t}}}
\end{equation}

where $n_{EPR}$ is the number of consumed EPR pairs, $n_{cov}$ is the number of covered gates, which may include more gates than those that were to be covered originally -- as shown in Fig.~\ref{fig:two_mig} -- , and $delay$ is the time, measured in discrete intervals, that must be waited before actually executing the gate. The $delay$ is estimated based on when the quantum links for entanglement generation were last used and when the gate should be executed. It may be the case that before executing a \texttt{TeleData} operation, one needs to wait for a previously scheduled one to complete. The $delay$ is scaled with the mean decoherence time $\bar{d_{t}}$ of the physical qubits.

\begin{algorithm}[ht!]
    \caption{Remote Gates Scheduler
        \newline
        \footnotesize
        \textbf{Input}: quantum circuit $QC$, network configuration $N$ and qubit assignment $P$
        \newline
        \textbf{Output}: quantum circuit with remote gates $D$
    }
    \label{alg:schedule}
    \begin{algorithmic}[1]
        \Function{Schedule}{}
            \State $D \gets \emptyset$
            \State $covered \gets \emptyset$
            \ForAll{$g \in QC$}
                \If{$g \notin D$}
                    \If{$g$ is local}
                        \State put $g$ into $covered$ and $D$
                    \Else
                        \State $TeleData \gets $ \textsc{Find TeleData($g$, $N$, $P$)}
                        \State $TeleGate \gets $ \textsc{Find TeleGate($g$, $N$, $P$)}
                        \If{\textsc{Cost($Teledata$) < \textsc{Cost($TeleGate$)}}}
                            \State put $TeleData$ into $D$
                        \Else
                            \State put $TeleGate$ into $D$
                        \EndIf
                        \State put $g$ into $covered$ and $D$
                        \State put extra covered gates into $D$ and $covered$
                    \EndIf
                \EndIf
            \EndFor
        \EndFunction
    \end{algorithmic}
\end{algorithm}

The pass selects the \texttt{TeleData} operation with the lowest cost and then covers a portion of the remaining uncovered gates\footnote{The dimension of the protion to cover is set by a customizable parameter.} with \texttt{TeleGate} operations. \texttt{TeleGate} operations are chosen and scheduled in a similar manner to the \texttt{TeleData} ones. \texttt{TeleGates} exploit the \texttt{Cat-Ent} primitive, as shown in Fig.~\ref{fig:tele_gate}~\cite{Yimsiriwattana2005}. Indeed, \texttt{TeleGate}s can be divided in three phases. First, with the \texttt{Cat-Ent} primitive, the one in the blue box in Fig.~\ref{fig:tele_gate}, the control qubit of a remote gate is ``shared'' with the QPU holding the target qubit. Then, gates controlled by the same shared controlled qubit can be be executed locally. Finally, with the \texttt{Cat-DisEnt} primitive, the one in the violet box in Fig.~\ref{fig:tele_gate}, the shared state of the control qubit is restored at the first QPU.

Both \texttt{TeleData} and \texttt{TeleGate} can either migrate one qubit to the other's QPU or both to a different one, as shown in Fig.~\ref{fig:migration}, depending on the cost of such operation (computed as in Eq.~\ref{eq:cover_cost}). Fig.~\ref{fig:one_mig} shows the first case, in which gate $g_0$ is covered by sharing qubit $q_1$ with $QPU_1$ through one \texttt{TeleGate} operation. Fig.~\ref{fig:two_mig} depicts the second case, where gate $g_0$ is covered by sharing qubits $q_1$ and $q_4$ with $QPU_1$, using two \texttt{TeleGate}s. By doing this, also gates $g_1$ and $g_2$ are covered. The same concept can be applied with \texttt{TeleData} operations.

The pass also compiles the same portion of circuit by scheduling only \texttt{TeleGate} operations. Finally, the pass can compute a cost for the portions of the circuit -- one with \texttt{TeleData} and \texttt{TeleGate}, the other with just \texttt{TeleGate} -- and select the strategy with the lowest cost. This time cost is simply the amount of consumed EPR pairs.
Finally, the pass resumes scanning gates in search of the next gate to cover.

\subsection{Local Routing}
\label{sec:routing}

The designed local routing pass takes as input a partitioned circuit with already scheduled remote operations and handles the local routing accordingly. The pass requires the partitioned circuit, with scheduled remote operations, the network configuration, and each QPU configuration, specifically coupling maps including connections between data qubits and communication qubits.

\begin{figure*}
    \centering
    \begin{minipage}[c]{.75\textwidth}
	\centering
        \includegraphics[width=1.\textwidth]{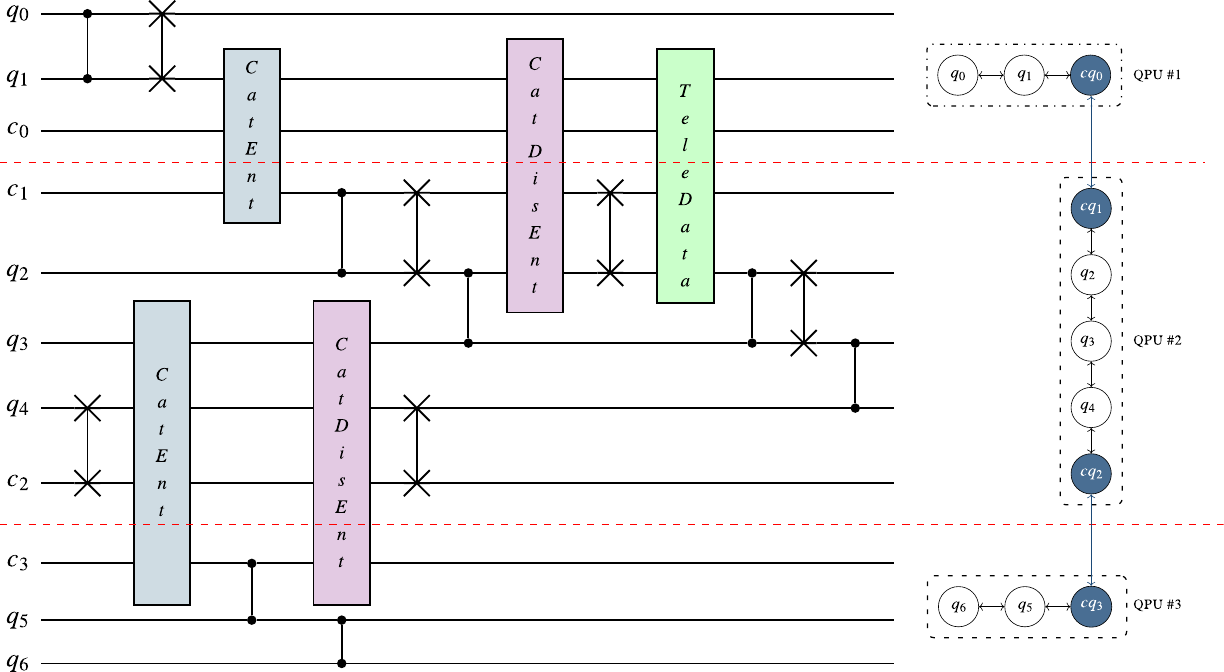}
    \end{minipage}
    \caption{Example of remote gate scheduling and local routing. Local gates are interlaced with \texttt{Cat-Ent}, \texttt{Cat-DisEnt} and \texttt{TeleData} operations as well as \texttt{SWAP} gates.}
    \label{fig:local_routing}
    \hrulefill
\end{figure*}

The core strategy is straightforward. The pass scans the circuit and for every gate that involves qubits not directly connected on their specific QPU, computes the shortest sequence of necessary \texttt{SWAP} gates. When it encounters a \texttt{TeleData} or \texttt{TeleGate} operation, it first checks if the involved data qubits are in proximity of one of the available communication qubits, among those corresponding to the quantum links used by the remote operation. If not, it computes the shortest paths to the less recently used communication qubit. The less recently used communication qubit is chosen to avoid as much delay as possible in entanglement generation.
At this stage of compilation, due to local \texttt{SWAP}s, the state of a data qubit may now reside on a communication qubit and vice versa. This is not necessarily an issue~\cite{Dahlberg2022}, but it is necessary to move the communication qubit back to its original position, after the remote operation is completed and before it is used again. This is crucial, not to loose the state of a data qubit physically stored at a communication qubit location, due to a new remote operation. An example of such instance in shown in Fig.~\ref{fig:local_routing}.

\section{Evaluation}
\label{sec:evaluation}
\begin{figure}[!h]
    \centering
    \includegraphics[width=.4\linewidth]{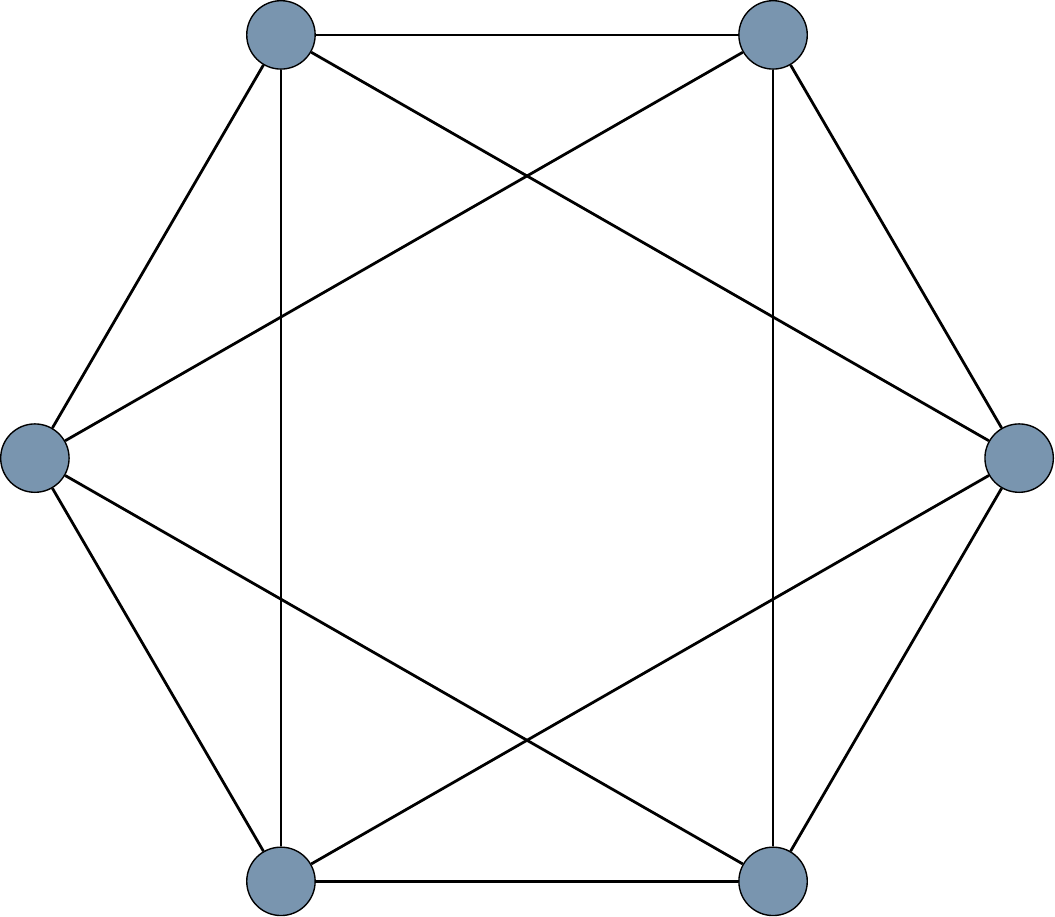}
    \caption{Example of graph state used to create graph state circuits. This graph state has 6 qubits, but can be scaled up to an arbitrary large number of qubits.}
    \label{fig:graph_state}
    \hrulefill
\end{figure}

\begin{figure}[!h]
    \centering
    \includegraphics[width=.7\linewidth]{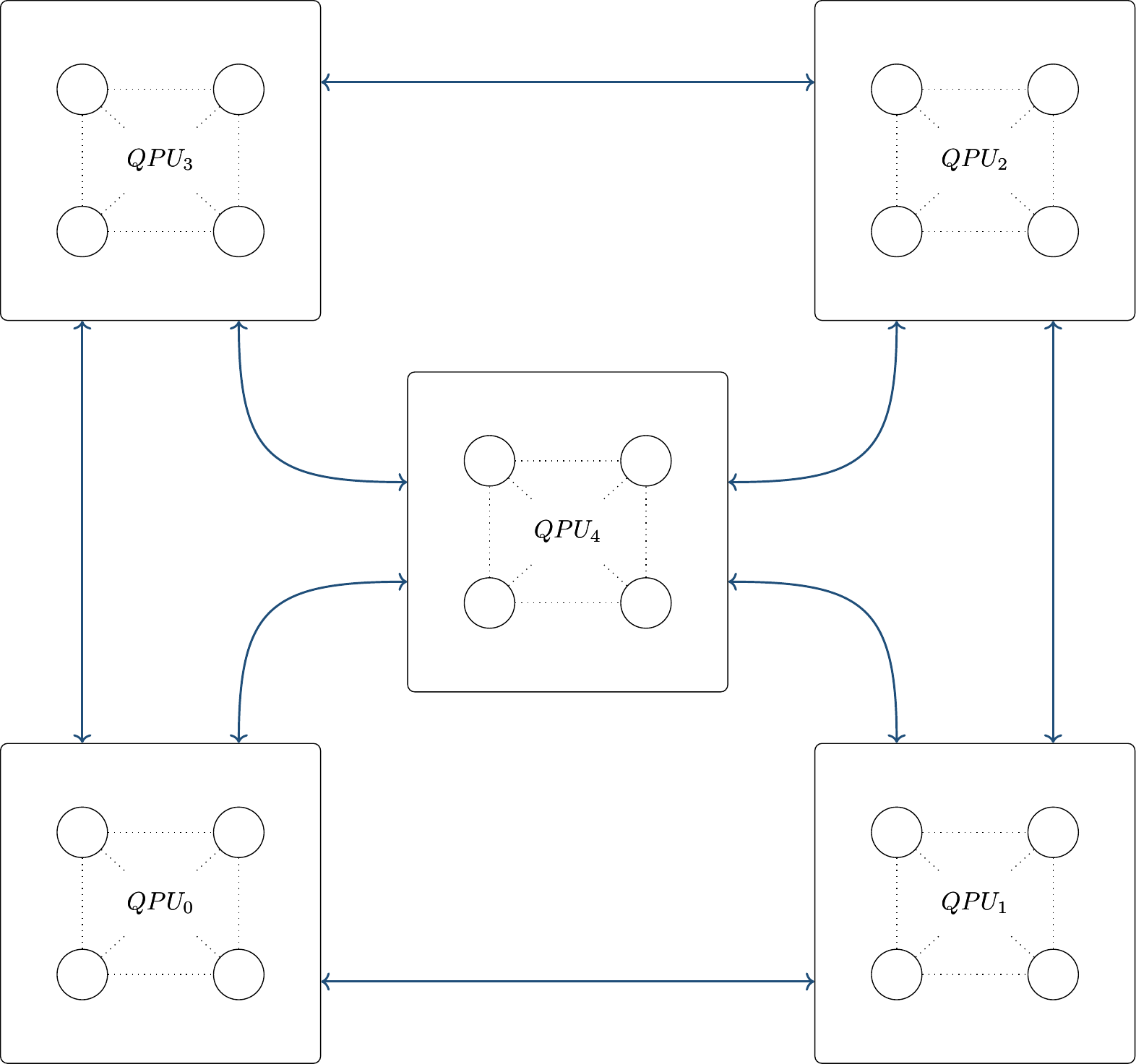}
    \caption{DQC architecture comprising 5 QPUs. Each QPU is connected to at least 3 others QPUs and, depending on the QPUs topology, there may be more than 1 \textit{communication qubit} per connection.}
    \label{fig:net_5}
    \hrulefill
\end{figure}

We implemented a quantum compiler based on the modular framework presented in Sec.~\ref{sec:compiler_framework}. The compiler was tested against three classes of quantum circuits, namely, VQE, QFT and Graph state circuits (an example of graph state used is shown in Fig.~\ref{fig:graph_state}). The aforementioned circuits were compiled for the DQC architecture illustrated in Fig.~\ref{fig:net_3} and Fig.~\ref{fig:net_5}, comprised respectively of 3 and 5 QPUs, denoted as \textit{Net-3} and \textit{Net-5}. To increase the number data qubits available, the QPUs in Fig.~\ref{fig:input_device} can be scaled up in a modular fashion~\cite{IBMTopologies}. In the following, \textit{QPU-n} denotes a QPU with \textit{n} data qubits. For each quantum circuit, the tests concerned remote gate scheduling with only the \texttt{TeleGate} operation, as well as with both the \texttt{TeleGate} and \texttt{TeleData} operations. For each compiled circuit the depth, the number of EPR pairs consumed and the layers dedicated to remote operations were used to analyze the results.

\begin{figure*}[!h]
    \centering
    \includegraphics[width=.9\textwidth]{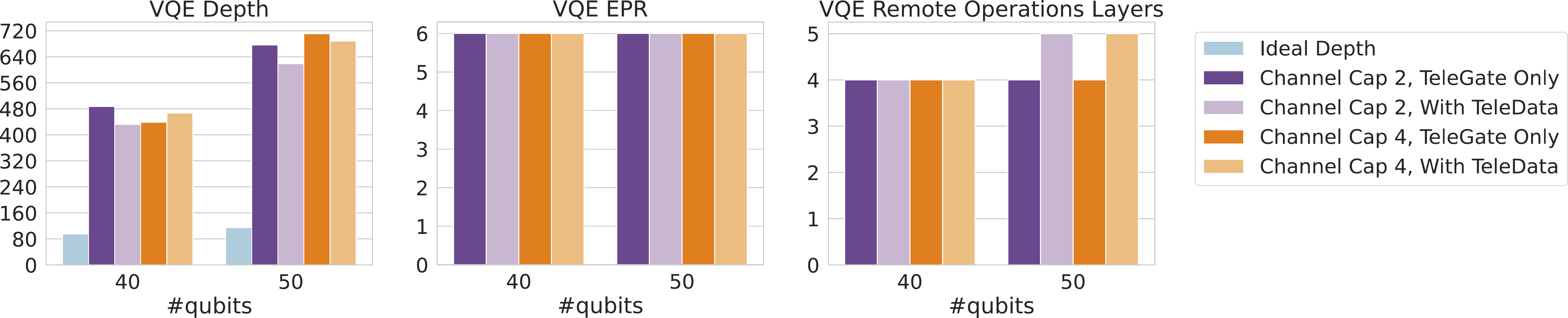}
    \caption{Results of VQE circuits compiled for the Net-3 with QPU-21. The number of qubits varies from 40 to 50, while the channel capacity varies from 2 to 4.}
    \label{fig:3QPU_21_vqe}
    \hrulefill
\end{figure*}

\begin{figure*}[!h]
    \centering
    \includegraphics[width=.9\textwidth]{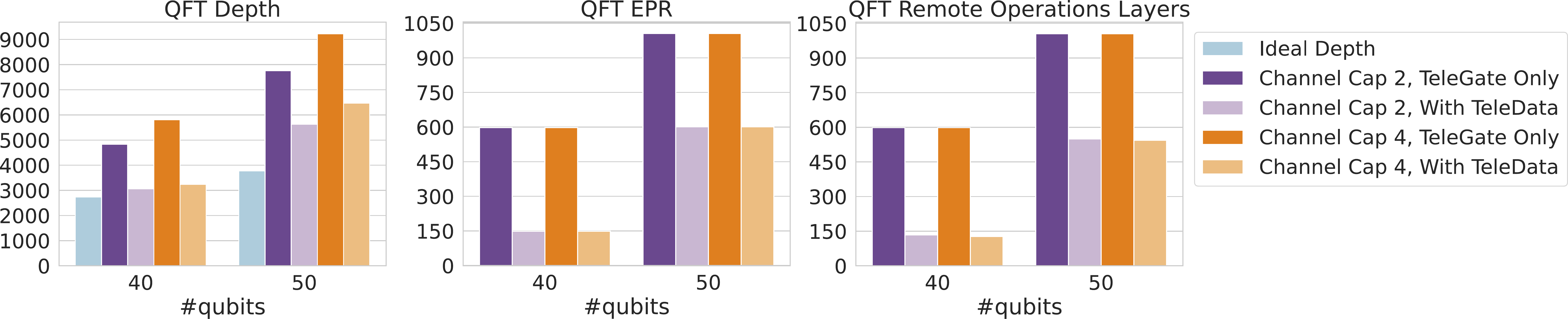}
    \caption{Results of QFT circuits compiled for the Net-3 architecture with QPU-21. The number of qubits varies from 40 to 50, while the channel capacity varies from 2 to 4.}
    \label{fig:3QPU_21_qft}
    \hrulefill
\end{figure*}

\begin{figure*}[!h]
    \centering
    \includegraphics[width=.9\textwidth]{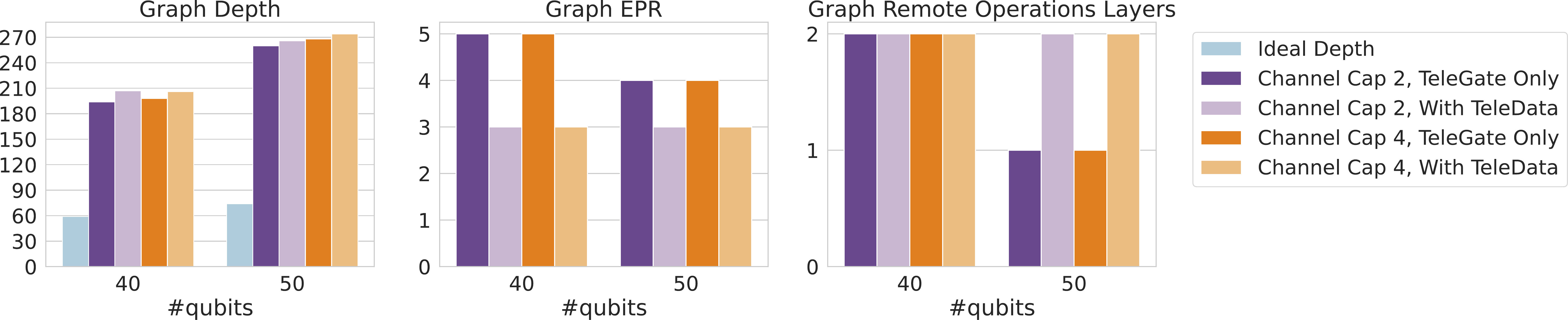}
    \caption{Results of Graph circuits compiled for the Net-3 architecture with QPU-21. The number of qubits varies from 40 to 50, while the channel capacity varies from 2 to 4.}
    \label{fig:3QPU_21_graph}
    \hrulefill
\end{figure*}

Fig.~\ref{fig:3QPU_21_vqe},~\ref{fig:3QPU_21_qft} and~\ref{fig:3QPU_21_graph} show compilation results of VQE, QFT and Graph circuits on Net-3 with QPU-21. It can be seen that exploiting \texttt{TeleData} operations alongside \texttt{TeleGate}s tends to be more beneficial, in terms of total depth, especially for QFT circuits. Moreover, using \texttt{TeleData} operations can greatly reduce the number of EPR pairs consumed and layers dedicated to remote operations for QFT circuits while being slightly detrimental for VQE circuits. Regarding Graph state circuits there is an increase in the remote operations layers, although minimal, which is opposed to a decrease in the number of EPR consumed, when using \texttt{TeleData} operations.

All figures show a slight increase in total circuit depth when the channel capacity changes from 2 to 4. At first glance, this may seem counter intuitive and it is probably an overhead caused by the local routing, which tries to use all available communication qubits, regardless of their distance from data qubits in the local coupling map.
Interestingly, there seems to be no difference in the number of layers dedicated to remote operations with respect to the channel capacity. We suppose that, due to the low connectivity between data qubits and communication qubits on each QPU, local routing operations create an upstream bottleneck with deleterious effects despite the increase in channel capacity. Further investigations in this regard will be necessary.

\begin{figure*}[!h]
    \centering
    \includegraphics[width=.9\textwidth]{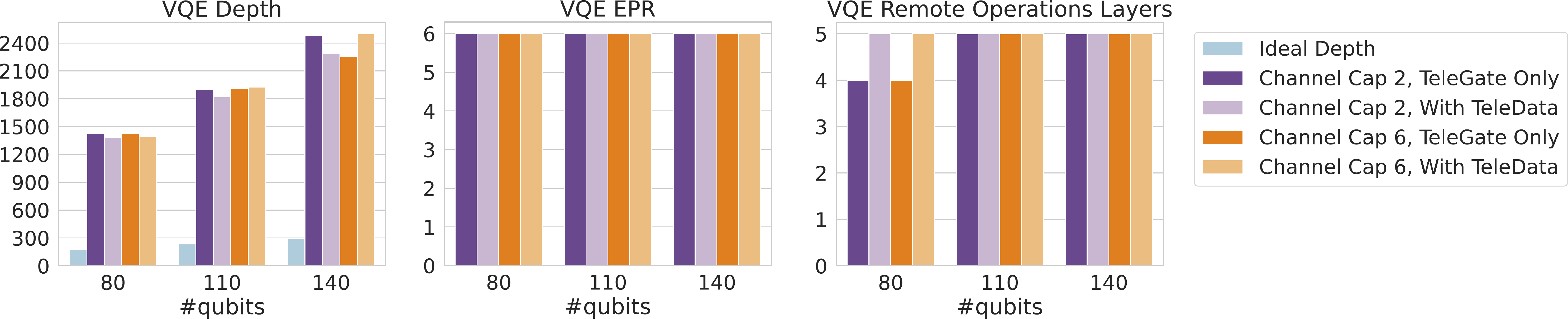}
    \caption{Results of VQE circuits compiled for the Net-3 with QPU-63. The number of qubits varies from 80 to 140, while the channel capacity varies from 2 to 6.}
    \label{fig:3QPU_63_vqe}
    \hrulefill
\end{figure*}

\begin{figure*}[!h]
    \centering
    \includegraphics[width=.9\textwidth]{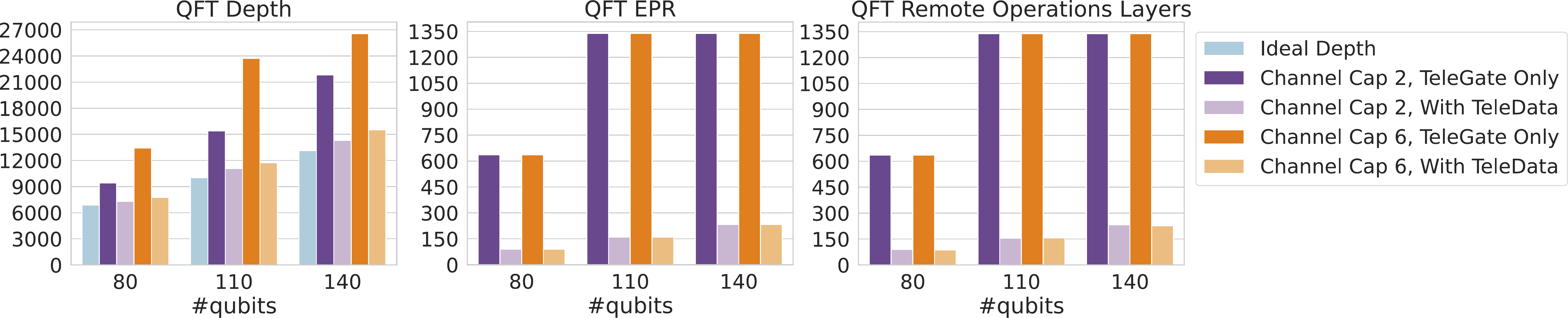}
    \caption{Results of QFT circuits compiled for the Net-3 architecture with QPU-63. The number of qubits varies from 80 to 140, while the channel capacity varies from 2 to 6.}
    \label{fig:3QPU_63_qft}
    \hrulefill
\end{figure*}

\begin{figure*}[!h]
    \centering
    \includegraphics[width=.9\textwidth]{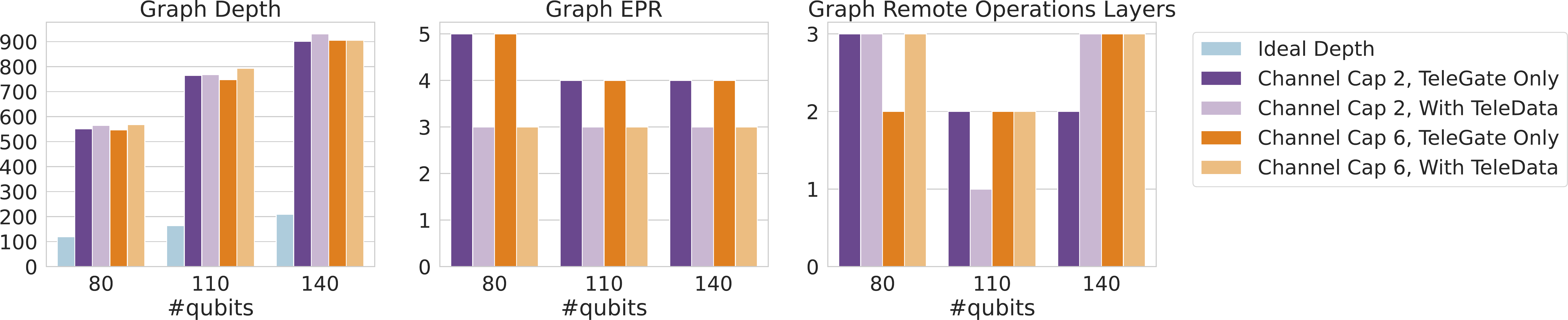}
    \caption{Results of Graph circuits compiled for the Net-3 architecture with QPU-63. The number of qubits varies from 80 to 140, while the channel capacity varies from 2 to 6.}
    \label{fig:3QPU_63_graph}
    \hrulefill
\end{figure*}

Some tests were also made using Net-3 with QPU-63 devices, with the number of data qubits used by the circuits ranging from 80 to 140. The results are reported in Fig.~\ref{fig:3QPU_63_vqe},~\ref{fig:3QPU_63_qft} and~\ref{fig:3QPU_63_graph}. While there is still not much of a difference when changing the channel capacity, the use of \texttt{TeleData} operations is greatly beneficial when distributing QFT circuits, which, from the number of EPR consumed, appear to be the circuit class that more heavily depends on remote operations, among those tested.

\begin{figure*}[!h]
    \centering
    \includegraphics[width=.9\textwidth]{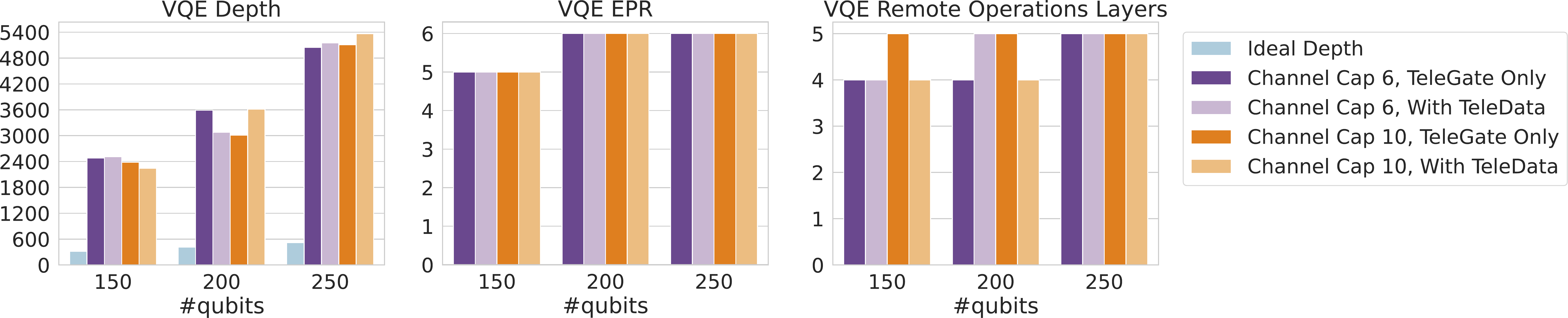}
    \caption{Results of VQE circuits compiled for the Net-3 with QPU-125. The number of qubits varies from 150 to 250, while the channel capacity varies from 6 to 10.}
    \label{fig:3QPU_125_vqe}
    \hrulefill
\end{figure*}

\begin{figure*}[!h]
    \centering
    \includegraphics[width=.9\textwidth]{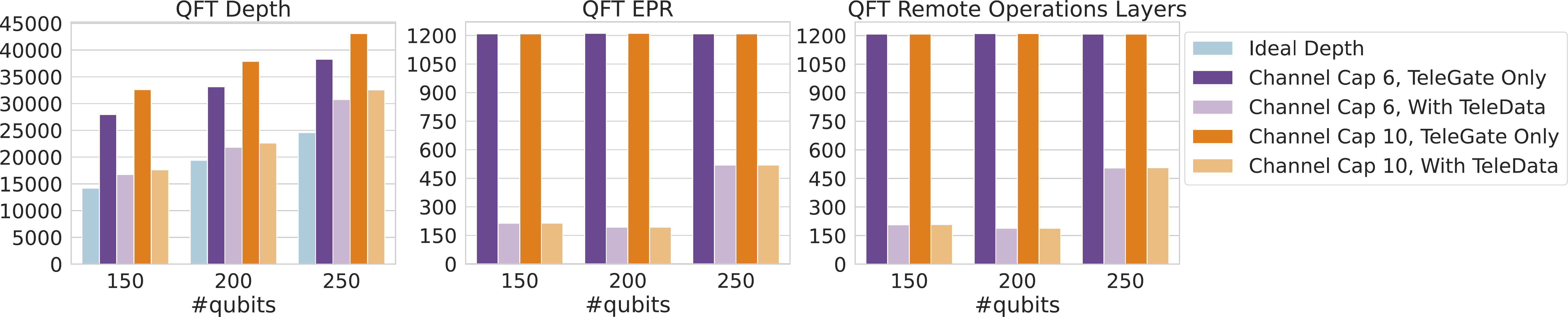}
    \caption{Results of QFT circuits compiled for the Net-3 architecture with QPU-125. The number of qubits varies from 150 to 250, while the channel capacity varies from 6 to 10.}
    \label{fig:3QPU_125_qft}
    \hrulefill
\end{figure*}

\begin{figure*}[!h]
    \centering
    \includegraphics[width=.9\textwidth]{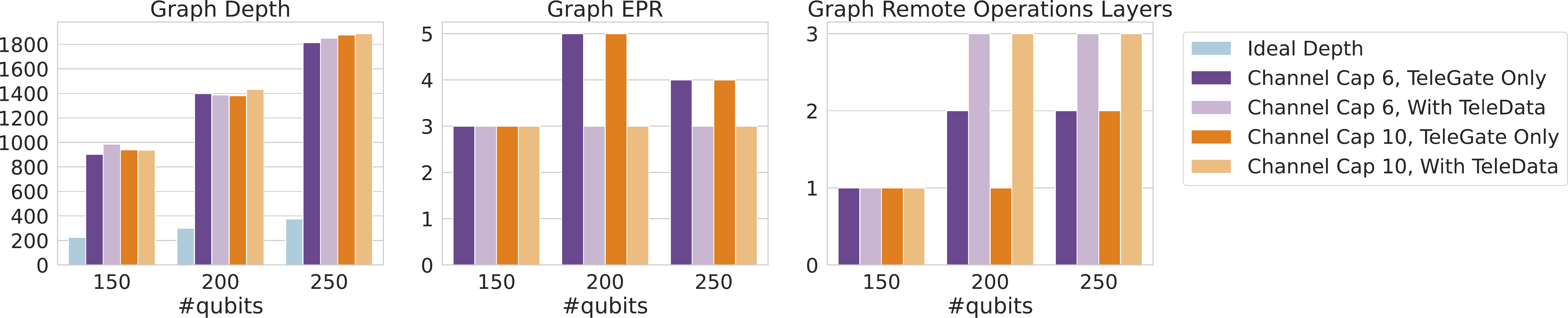}
    \caption{Results of Graph circuits compiled for the Net-3 architecture with QPU-125. The number of qubits varies from 150 to 250, while the channel capacity varies from 6 to 10.}
    \label{fig:3QPU_125_graph}
    \hrulefill
\end{figure*}

By maintaining Net-3 but changing to QPU-125 devices, the compiler was tested on circuits with up to 250 qubits. At this stage, an interesting observation can be made from Fig.~\ref{fig:3QPU_125_vqe},~\ref{fig:3QPU_125_qft} and~\ref{fig:3QPU_125_graph}. It seems that, for Graph states circuits, when the number of qubits grows, and the data qubits capacity of the network is topped up, while the number of EPR pairs consumed remains unchanged, there is an almost unnoticeable increase in the number of layers for remote operations.

\begin{figure*}[!h]
    \centering
    \includegraphics[width=.9\textwidth]{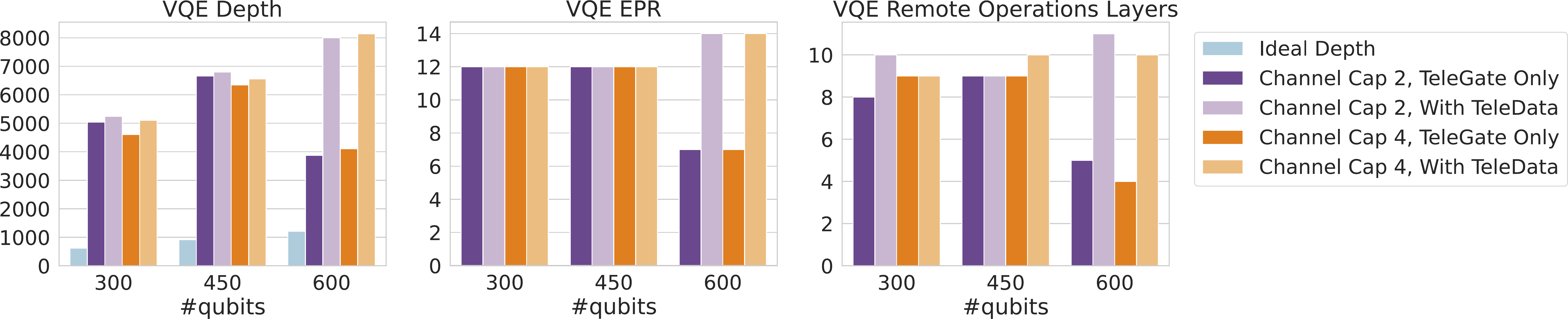}
    \caption{Results of VQE circuits compiled for the Net-5 with QPU-125. The number of qubits varies from 300 to 600, while the channel capacity varies from 2 to 4.}
    \label{fig:5QPU_125_vqe}
    \hrulefill
\end{figure*}

\begin{figure*}[!h]
    \centering
    \includegraphics[width=.9\textwidth]{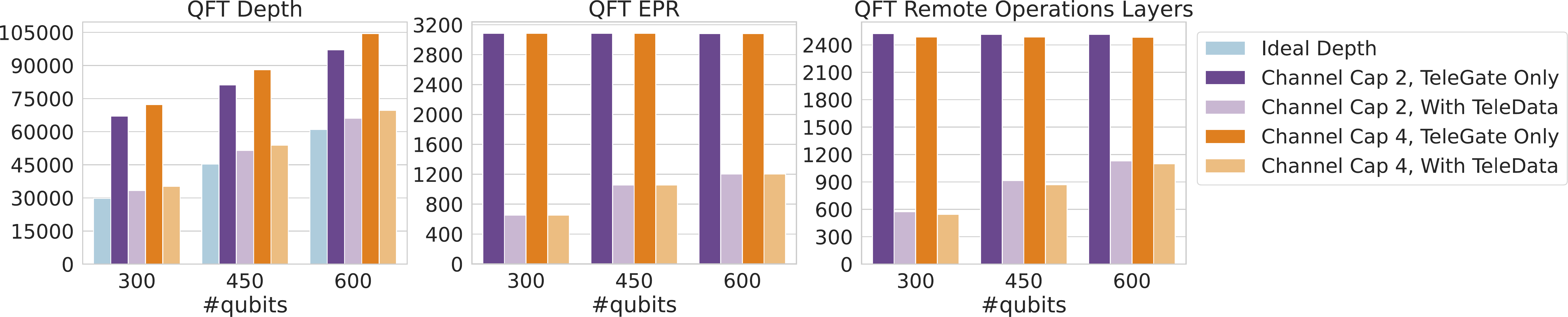}
    \caption{Results of QFT circuits compiled for the Net-5 architecture with QPU-125. The number of qubits varies from 300 to 600, while the channel capacity varies from 2 to 4.}
    \label{fig:5QPU_125_qft}
    \hrulefill
\end{figure*}

\begin{figure*}[!h]
    \centering
    \includegraphics[width=.9\textwidth]{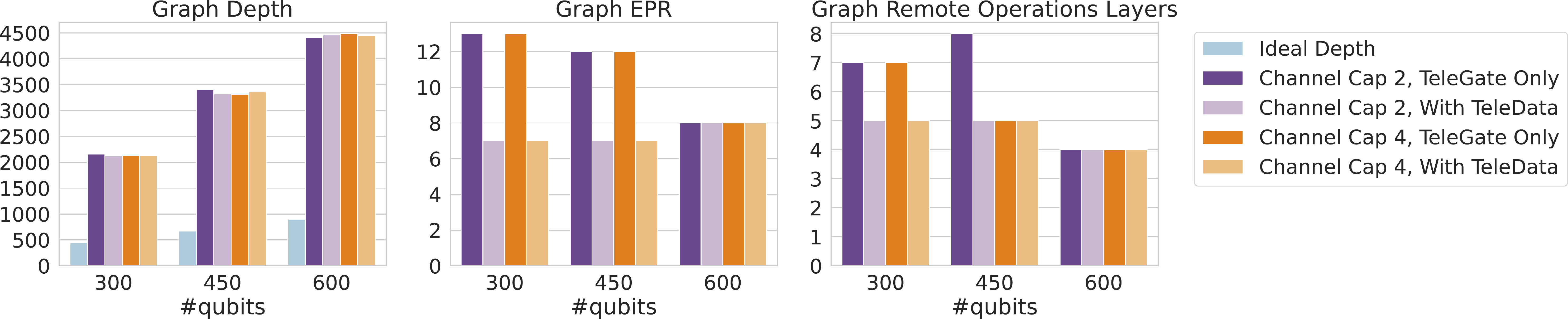}
    \caption{Results of Graph circuits compiled for the Net-5 architecture with QPU-125. The number of qubits varies from 300 to 600, while the channel capacity varies from 2 to 4.}
    \label{fig:5QPU_125_graph}
    \hrulefill
\end{figure*}

Finally, the total number of data qubits was further increased, by exploiting Net-5 with QPU-125 devices. Therefore, it was possible to compile circuits up to 600 qubits, as depicted in Fig.~\ref{fig:5QPU_125_vqe},~\ref{fig:5QPU_125_qft} and~\ref{fig:5QPU_125_graph}. There are two results that stand out in these figures. Firstly, for VQE circuits, the results show that there is a slight increase in the layers of remote operations when the maximum number of qubits is reached and \texttt{TeleData} operations are employed. The opposite can be observed for Graph state circuits, where the number of remote operations layers decreases, although marginally, when the maximum number of data qubits allowed by the network is reached. This trend goes against the observation made previously for the same type of circuits albeit with the Net-3 topology, which outlines the impact of different network topologies and suggests that choosing a more connected network is in fact beneficial.

\section{Conclusion}
\label{sec:conclusion}
In this work, we introduced a general-purpose modular quantum compilation framework for DQC that takes into account both network and device constraints and characteristics.
We illustrated the experimental evaluation of a quantum compiler based on the proposed framework, using some circuits of interest (VQE, QFT, graph state preparation) characterized by different widths (up to 600 qubits). We considered different network topologies, with quantum processors characterized by heavy hexagon coupling maps. We also presented a strategy for remote scheduling that can exploit both \texttt{TeleGate} and \texttt{TeleData} operations, and tested the impact of using either only \texttt{TeleGate}s or both operations. We observed that \texttt{TeleData} operations may have a positive impact on the number of consumed EPR pairs. Furthermore, we showed that choosing a more connected network topology helps reduce the number of layers dedicated to remote operations.

Regarding future work, we will focus on integrating noise-adaptive compilation strategies into the framework, both for local routing~\cite{FerAmo2022} and remote gate scheduling. We shall then evaluate the impact of different strategies on the quality of computation results, which depend also on the selection of suitable metrics. To produce such metrics we need to actually execute the compiled circuits, either by means of a quantum network simulator or on real hardware. In the first case, there are already available simulators with different levels of abstraction, depending on how realistic the simulations needs to be. These simulations will be crucial to understand the impact that remote operations, and any resulting local routing overhead, have on the quality of the computation due to the effects of noise.

\section*{Acknowledgement}
The authors acknowledge financial support from the EU Flagship on Quantum Technologies through the project Quantum Internet Alliance (EU Horizon Europe, grant agreement no. 101102140).
This research benefits from the HPC (High Performance Computing) facility of the University of Parma, Italy.

\subsection*{Data Availability}
All data and code required to reproduce all plots shown herein are available at \url{https://doi.org/10.5281/zenodo.7896589}.

\bibliographystyle{IEEEtran}
\bibliography{bibliography.bib}


\begin{small}

\begin{wrapfigure}{l}{25mm} 
    \includegraphics[width=1in,height=1.25in,clip,keepaspectratio]{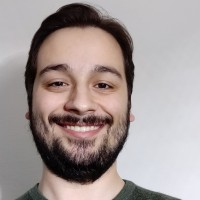}
  \end{wrapfigure}\par
  \textbf{Davide Ferrari} (GSM'20) received the MSc degree in Computer Engineering from the University of Parma, Italy, in 2019. Right after, he has been a research scholar at Future Technology Lab of the University of Parma, working on the design of efficient algorithms for quantum compiling. During his Ph.D. in Information Technologies at the Department of Engineering and Architecture of the University of Parma, he worked on quantum compiling, quantum optimization and distributed quantum computing. He is now a research fellow at the Department of Engineering and Architecture of the University of Parma. He is involved in the Quantum Information Science (QIS) research initiative at the University of Parma, where he is a member of the Quantum Software research unit. In 2020, he won the 'IBM Quantum Awards Circuit Optimization Developer Challenge'. His research focuses on quantum optimization applications and efficient quantum compiling for local and distributed quantum computing.\par

\vspace{0.2cm}

\begin{wrapfigure}{l}{25mm} 
    \includegraphics[width=1in,height=1.25in,clip,keepaspectratio]{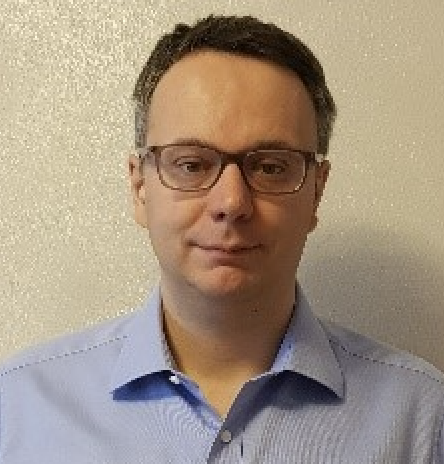}
  \end{wrapfigure}\par
  \textbf{Stefano Carretta} received his PhD in Physics from the University of Parma (Italy) in 2005 and he is Full Professor in Physics of Matter at the University of Parma. His research activity is mainly focused on the theoretical modelling of the quantum behavior of magnetic molecules and in quantum information processing. He contributed to put forward some of the first proposals for the use of magnetic molecules as qubits and the first proposal for exploiting molecular nanomagnets as quantum simulators. He was member of the commission of experts on Quantum Technologies for the 2021-27 Italian National Research Program (PNR) and he is involved into several national and European projects involving Quantum Technologies. He is currently one of the PI of an European ERC Synergy Project. He is coauthor of more than 150 research papers published in international journals. In 2006 he has been appointed of the "Le Scienze" (the Italian version of Scientific American) medal and of the President of the Italian Republic medal for his research on molecular nanomagnets. In 2011 he has been appointed of the prestigious Olivier Kahn International Award for his contribution to the theory of molecular magnetism.\par

\newpage

\begin{wrapfigure}{l}{25mm} 
    \includegraphics[width=1in,height=1.25in,clip,keepaspectratio]{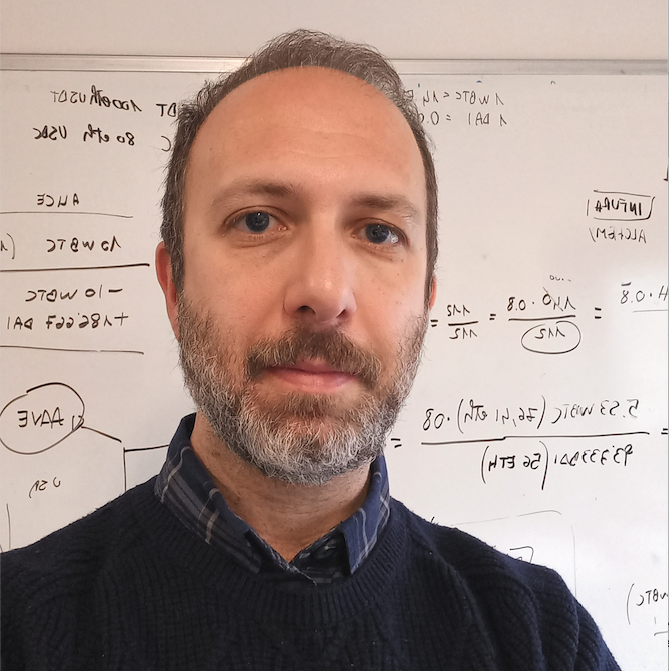}
  \end{wrapfigure}\par
  \textbf{Michele Amoretti} (S'01-M'06-SM'19) received his PhD in Information Technologies in 2006 from the University of Parma, Parma, Italy. He is Associate Professor of Computer Engineering at the University of Parma. In 2013, he was a Visiting Researcher at LIG Lab, in Grenoble, France. His current research interests are mainly in Quantum Computing, High Performance Computing and the Internet of Things. He authored or co-authored over 130 research papers in refereed international journals, conference proceedings, and books. He serves as \textit{Associate Editor} for the journals: IEEE Trans. on Quantum Engineering and International Journal of Distributed Sensor Networks. He is involved in the Quantum Information Science (QIS) research and teaching initiative at the University of Parma, where he leads the Quantum Software Laboratory. He is currently one of the PI of the European HE project Quantum Internet Alliance. He is the CINI Consortium delegate in the UNI/CT 535 ``Quantum Technologies'' UNINFO Commission, which is the national mirror of ISO/IEC JTC 1 WG 14 ``Quantum Information Technologies'' and of CEN/CENELEC JTC 22 ``Quantum Technologies''.\par

\end{small}

\end{document}